\documentclass[11pt,twoside]{article}

\usepackage{asp2006}
\usepackage{epsf}

\begin{document}
\title{HI Cosmology in the Local Universe with ALFALFA}   
\author{Martha P. Haynes}   
\affil{Center for Radiophysics and Space Research,
Cornell University, Ithaca, NY 14853 USA}    

\begin{abstract} 
The Arecibo Legacy Fast ALFA (ALFALFA) survey is an on-going second
generation blind extragalactic HI survey exploiting Arecibo's superior 
sensitivity, angular resolution and digital technology to conduct a
census of the local HI universe over a cosmologically significant volume.
As of mid-2007, $\sim$4500 good quality extragalactic
HI line sources have been extracted in $\sim$15\% of the final survey area.
ALFALFA is detecting HI masses as low as 10$^6 M_\odot$~and
as high as 10$^{10.8} M_\odot$~with positional accuracies typically better
than 20\arcsec, allowing immediate identification of the most probable
optical counterparts. Only 3\% of all extragalactic HI sources
and fewer than 1\% of detections with
$M_{HI} > 10^{9.5} M_\odot$ cannot be identified with a stellar component.
First ALFALFA results already suggest, in agreement with previous studies,
that there does not appear to be a cosmologically significant population of
optically dark but HI rich galaxies.  ALFALFA promises
a wealthy dataset for the exploration of many issues in near-field cosmology
and galaxy evolution studies, setting the stage for their extension
to higher redshifts with the Square Kilometer Array (SKA).
\end{abstract}



\section{Introduction}

21~cm HI line observations 
over the years have provided critical understanding of the nature of galaxies
and the processes which govern their evolution. HI line
studies address a host of fundamental cosmological questions (the number density, 
distribution and nature of optically--dark and low mass halos) 
and issues of galaxy formation and
evolution (sizes of HI disks, history of tidal interactions and mergers,
low z absorber cross section, origin of dwarf galaxies, nature of high velocity 
clouds). The program for the current symposium {\it ``Frontiers of Astrophysics''},
marking the 50th anniversary of NRAO, illustrates the continued promise of
HI line investigations not only for the exploration of galaxies 
and large scale structure but also for new applications which will 
explore entirely uncharted cosmic volumes, the ``Dark Ages''. 
Surveys of redshifted HI for cosmological purposes are main drivers of the 
Square Kilometer Array (SKA). 

However, it is important to keep in mind
that, in comparison with the view gleaned from optical redshift catalogs, 
our understanding of $z=0$ cosmology through the eyes of the 21~cm line is
still immature. The HI equivalent of the optical luminosity function (OLF),
the HI Mass Function (HIMF), is still derived from only a few thousand 
objects. Among recent estimates of the HIMF are those
presented by Zwaan {\it et al} (1997), Rosenberg \& Schneider (2002), 
Zwaan {\it et al} (2003) and Springob {\it et al} (2005). The latter is
based on a compilation of HI observations of $\sim$9000 optically selected 
galaxies, further restricted by HI line flux and optical diameter to a complete subsample
containing 2200 objects. The other determinations are based on blind
HI surveys and thus have no bias against the low luminosity and low
surface brightness galaxies which may be underrepresented in
optical galaxy catalogs. However, all of these studies suffer from limitations
associated with small number statistics, systematics and the effects of cosmic 
variance, especially at both the low and high mass ends. Of particular note, 
the lowest mass objects are detected only very nearby; determination of the 
low mass HIFM slope is not only limited by small number statistics but also
by large uncertainties in the HI masses derived using redshift 
distances (Masters, Haynes \& Giovanelli 2004). Because of volume limitations,
the  high mass end is likewise poorly sampled. Since the most massive objects 
will be the HI lampposts detected at high redshift in large numbers by the 
SKA and its precursor instruments, it is critical to understand the
nature and distribution of their local counterparts.

Here, I introduce and review the promise of the on-going
Arecibo Legacy Fast ALFA (ALFALFA) survey to yield cosmologically
significant results on the number and distribution
of optically-dark and low mass halos and on the determination 
of the $z = 0$ HIMF, the HI correlation function and its bias parameter.

\section{ALFALFA: A Second Generation Blind HI Survey}

Just as the introduction of wide field CCDs revolutionized the survey capabilities
of optical and infrared telescopes, HI line astronomy is undergoing a similar 
renaissance with the advent of multi-beam receivers on the large single-dish 
telescopes, enabling blind HI surveys that cover wide areas. Table 
\ref{tab:haynes_tab1} summarizes the principal characteristics of major blind 
HI surveys. Most recently, the 305~m antenna has been equipped with a 7-beam
system known as ALFA, the Arecibo L-band Feed Array, which is being used
to conduct wide area surveys in galactic, extragalactic and pulsar research.
The local extragalactic sky visible to 
Arecibo is rich, containing the central longitudes of the Supergalactic Plane 
in and around the Virgo cluster, the main ridge of the Pisces--Perseus 
Supercluster, and the extensive filaments connecting A1367, Coma and Hercules. 
Included in Table \ref{tab:haynes_tab1}
are two surveys currently exploiting ALFA, namely ALFALFA and a somewhat deeper
targeted survey covering a much smaller area, the Arecibo Galaxy Environments
Survey (AGES; Auld {\it et al} 2006). 

\begin{table}[!t]
\caption{Comparison of Major Blind HI Surveys}
\smallskip
\begin{center} 
\begin{tabular}{ccrccrrc}
\hline
\noalign{\smallskip}
Survey & Beam      & Area   & $\delta V$    & rms$^a$ & V$_{median}$ & N$_{det}$ & Ref\\
       & ($^{\prime}$) & ($deg^2$) & (km~s$^{-1}$) & ($^a$)  & (km~s$^{-1}$) &     &    \\
\noalign{\smallskip}
\hline
\noalign{\smallskip}
AHISS   & 3.3 &    13 & 16 & 0.7 & 4800 &   65 &  $^b$ \\
ADBS    & 3.3 &   430 & 34 & 3.3 & 3300 &  265 &  $^c$ \\
WSRT    & 49. &  1800 & 17 & 18  & 4000 &  155 &  $^d$ \\
HIPASS  & 15. & 30000 & 18 & 13  & 2800 & 5000 &  $^{e,f}$ \\
HI-ZOA  & 15. &  1840 & 18 & 13  & 2800 &  110 &  $^g$ \\
HIDEEP  & 15. &    32 & 18 & 3.2 & 5000 &  129 &  $^h$\\
HIJASS  & 12. &  1115 & 18 & 13  & $^i$ &  222 &  $^i$\\
J-Virgo & 12. &    32 & 18 &  4  & 1900 &   31 &  $^j$\\
AGES    & 3.5 &   200 & 11 & 0.7 & 12000&  $\dots$    &  $^k$\\
ALFALFA & 3.5 &  7000 & 11 & 1.7 & 7800 &$>$25000& $^l$\\
\noalign{\smallskip}
\hline
\end{tabular}\label{tab:haynes_tab1}

{\small
$^a$ mJy per beam at 18 km~s$^{-1}$ resolution;
$^b$ Zwaan {\it et al} (1997);
$^c$ Rosenberg \& Schneider (2000);
$^d$ Braun, Thilker \& Walterbos (2003);
$^e$ Meyer {\it et al} (2004);
$^f$ Wong {\it et al} (2006);
$^g$ Henning {\it et al} (2000);
$^h$ Minchin {\it et al} (2003);
$^i$ Lang {\it et al} (2003), HIJASS has a 
gap in velocity coverage between 4500-7500 km~s$^{-1}$, caused by RFI;
$^j$ Davies {\it et al} (2004);
$^k$ Auld {\it et al} (2006);
$^l$ Giovanelli {\it et al} (2007).}
\end{center}
\end{table}

ALFALFA is a two-pass, fixed azimuth spectral line survey which aims 
to map 7000 deg$^{2}$ of high galactic latitude sky over the HI velocity
range from $-1600$ to $+18000$ km~s$^{-1}$.
Exploiting the big dish's large collecting area and small beam size, ALFALFA is 
specifically designed to probe the faint end of the HI mass function, a goal 
for which sky coverage is critical (Giovanelli
{\it et al} 2005a). It exploits a ``minimum intrusion'' technique designed
to maximize data quality (Giovanelli {\it et al} 2005b) and an automated 
Fourier domain signal extraction algorithm (Saintonge 2007). ALFALFA
has $8\times$ the sensitivity, $4\times$  the angular resolution, 
$3\times$ the  spectral resolution, and $1.6\times$ the total bandwidth of 
HIPASS, the HI Parkes All-Sky Survey (Zwaan {\it et al} 2004; 
Meyer {\it et al} 2004). With a median redshift of only $\sim$2800 km~s$^{-1}$,
HIPASS did not sample adequate extragalactic volume to yield a
cosmologically ``fair sample''. Furthermore, the large beamsize
(15\arcmin) of the Parkes telescope made identification of optical
counterparts often uncertain without followup HI synthesis observations
(Oosterloo {\it et al} 2007). The centroiding accuracy of ALFALFA is
on average 24\arcsec~ (20\arcsec~ median) for all sources with
signal-to-noise ratio $> 6.5$ so that 
identification of the most probably stellar counterpart is an integral
part of the ALFALFA cataloging process. Its median redshift is
$\sim$7800 km~s$^{-1}$, nearly $3\times$ that of HIPASS.

Initial results from ALFALFA precursor observations were published
by Giovanelli {\it et al} (2005b), and two catalogs of HI detections
from the survey itself are published (Giovanelli {\it et al} 2007)
or submitted (Saintonge {\it et al} 2007). Catalogs, spectra and associated
data are
being made available from the Cornell Digital HI Archive 
{\it http://arecibo.tc.cornell.edu/hiarchive/}. The first catalog
(Giovanelli {\it et al} 2007) covering $11^h44^m <$ R.A. $< 14^h00^m$,
 $+12^\circ < $ Decl. $<+16^\circ$, contains 730 
HI detections and their most probable optical counterparts. 
In comparison, HIPASS detected only 40 HI sources in the same region, two of 
which are unconfirmed by ALFALFA. 
Although this region of the sky has been heavily surveyed by previous targeted
observations based on optical flux-- or size-- limited
samples, 69\% of the extracted sources are newly reported HI detections, an
indication that previous criteria for identifying potentially
HI--rich targets have neglected most of the HI-rich population.

\begin{figure}[!t]
\plotone{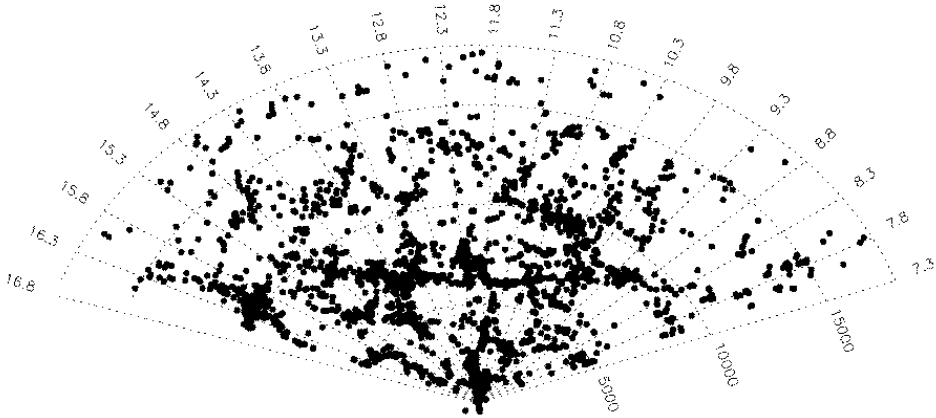}
\caption{Cone plot 2657 HI sources detected by ALFALFA in the region R.A.=[7.5$^h$--16.5$^h$],
Dec=[12$^\circ$--16$^\circ$], which represents 7.5\% of the survey. Note that due
to RFI, ALFALFA is effectively blind in the redshift range
between approximately 15000 and 16000 km~s$^{-1}$.}\label{fig:haynesfig1}
\end{figure}

\begin{figure}[!t]
\plotone{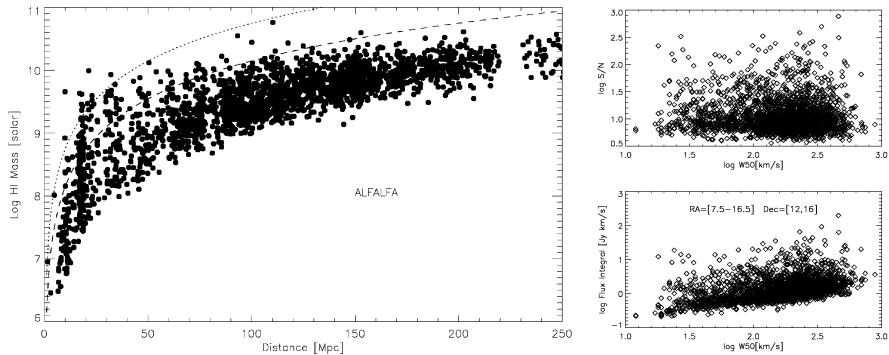}
\caption{{\bf Left:} Sp\" anhauer diagram of the 2657 HI sources included in
Figure \ref{fig:haynesfig1}. The two smooth lines identify respectively the 
completeness limit
(dotted) and the detection limit (dashed) for sources of 200 km~s$^{-1}$ linewidth for
the HIPASS survey. Note that due to RFI, ALFALFA is
effectively blind in the redshift range between approximately 15000 and 16000 km~s$^{-1}$.
{\bf Right:} Signal--to--noise ratio versus velocity width (top) and integrated HI line flux
density versus velocity width for the same galaxies.
}\label{fig:haynesfig2}
\end{figure}

At the time of this meeting, signal extraction has been completed for $\sim$15\% 
of the survey area with catalogs totaling $\sim$4500 detections in preparation for publication
in the second half of 2007. The largest contiguous region
fully processed to date corresponds to a strip between 7.5$^h <$ R.A. 
$< 16.5^h$, 12$^\circ <$ Decl. $< 16^\circ$. A total of 2657 HI sources have
been detected in that region which covers $\sim$7.5\% of the survey total
solid angle. Figure \ref{fig:haynesfig1} shows a cone diagram of the
ALFALFA detections in this strip; the distribution of HI sources
closely matches the complex features of large scale structure evident also
from optical surveys. Figure \ref{fig:haynesfig2}
summarizes the characteristics of the HI detections in the same 
region. The statistical improvement of ALFALFA over HIPASS is clear from
the Sp\" anhauer diagram shown on the left; in fact, in the same region,
HIPASS detected only 90 sources versus the 2700 found by ALFALFA. 
The diagrams on the right
show the quality of the ALFALFA data and signal extraction technique:
the S/N of detections exhibits no significant bias with respect to velocity width.
Spectroscopic HI surveys are not simply flux limited; the flux limit is expected
to increase as $W^{1/2}$ for low velocity widths and show a linear rise for the
wider line profiles. Such a transition is observed near $\log W\simeq 2.5$. The
ALFALFA flux limit is $\sim 0.25$ Jy~km~s$^{-1}$ for narrow lines, 
rising to 1 Jy~km~s$^{-1}$ for the broadest ones.
The detection areal density of $\sim$5 sources deg$^{-2}$, with peaks $10-20\times$
higher in regions of groups and clusters
suggests that the full ALFALFA survey may catalog as many as
30,000 HI sources, a higher yield than suggested by survey
simulations based on estimates of the HIMF derived from 
previous surveys (Giovanelli {\it et al} 2005a). 

\section{ALFALFA and the Existence of Massive ``Dark Galaxies''}\label{sec:haynes_dark}

A truly ``dark galaxy'' is a halo consisting only of dark matter. 
In some scenarios, it is possible that some optically dark objects
may contain enough HI that a blind HI survey would detect
them. A good example of a ``dark'' object is the southwestern component
of the binary system known as HI1225+01. While the northeastern HI component hosts a small,
star forming dwarf, the SW component has no detectable stellar counterpart. VLA observations reveal a velocity
field implying rotation with $V_{rot}\simeq 14$ km~s$^{-1}$, yielding a dynamical mass of $10^9$ M$_\odot$ 
and thus $M_{dyn}/L>200$ (Chengalur, Giovanelli \& Haynes 1995). However, it is not
an isolated object, being part of an apparent binary system.

\begin{figure}[!t]
\plotone{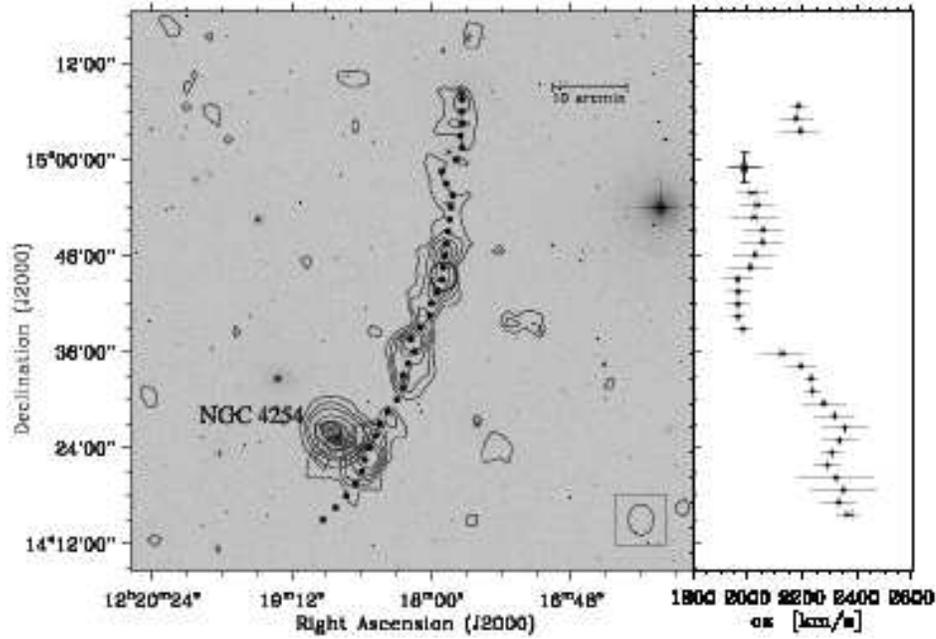}
\caption{Left: HI flux contours extracted from the ALFALFA
survey dataset, superposed on a DSS2 Blue image.  The 36 filled dots
indicate the locations of successive deeper L-band wide (LBW) receiver
pointings (Haynes {\it et al} 2007).  The contours centered on NGC~4254 are at 10, 15, 20, 30, and 40
 Jy beam$^{-1}$ km~s$^{-1}$, integrated from 2259 to 2621 km~s$^{-1}$.
The contours for the HI stream are at 0.35, 0.52, 0.70, 0.87, and 1.0 Jy beam$^{-1}$ km~s$^{-1}$,
integrated from 1946 to 2259 km~s$^{-1}$.
The 3\arcmin ~circle in mid stream indicates the position of VirgoHI21 (Minchin {\it et al} 2007). 
Right: The velocity of the HI emission peak 
as seen in LBW pointings.}\label{fig:haynesfig3}
\end{figure}

A candidate isolated massive dark galaxy VirgoHI21, was detected by 
Davies {\it et al} (2004) during the HI Jodrell All-Sky Survey (HIJASS) and 
corroborated by Arecibo and WSRT observations (Minchin {\it et al} 2007).
VirgoHI21 lies some 100 kpc N of NGC~4254, M99, one of the brightest spiral
galaxies in the Virgo Cluster. Well-known for its prominent optical lopsidedness
and strong $m=1$ spiral mode, NGC~4254 lies about 3$^\circ$, or 1 Mpc at 
the Virgo distance, from M87 and appears to be moving at high speed relative to the
cluster. It is, however, not HI deficient. As illustrated in Figure
\ref{fig:haynesfig3}, the ALFALFA observations reported by Haynes {\it et al} (2007)
show clearly the existence of a huge HI stream, in which VirgoHI21 is one
condensation, extending $\sim$ 250 kpc
to the north of NGC~4254. The gas mass associated with the 
stream is a modest $6 \times 10^8$ $M_\odot$, or about 10\% of the total 
HI associated with NGC~4254. One of the driving arguments
for the interpretation of VirgoHI21 as an isolated massive dark 
galaxy is the gradient seen in the velocity field (Minchin {\it et al}
2007); as evident in Figure \ref{fig:haynesfig3},
this gradient is just part of the varying, large--scale velocity field
along the stream.
Duc \& Bournaud (2007) have modeled the formation of a tidal
tail closely resembling that seen in Figure \ref{fig:haynesfig3}
by a a very high speed encounter of NGC~4254 with another cluster
member now located several degrees away. ALFALFA has discovered a number of 
other optically-dark HI clouds in the periphery of Virgo (Kent
{\it et al} 2007; Kent, this volume) which may be of similar origin.

One of the principal results of HIPASS is the lack of
HI detections without optical counterparts in those regions of
the sky not obscured by the Galactic plane. ALFALFA will deliver much
more stringent constraints on the existence of gas-rich dark galaxies
through its sensitive, large volume coverage.

\section{ALFALFA and the ``Missing Satellite Problem''}\label{sec:haynes_missing}

One of the principal discrepancies between cold dark 
matter theory and current observations revolves around the large difference 
between the number of dwarf dark matter halos seen around giant halos in 
numerical simulations based on CDM and the observed dwarf satellite population 
in the Local Group (Kauffmann {\it et al} 1993; Klypin {\it et al} 1999), referred
to as the ``missing satellite problem''.
Previous determinations of the HIMF below 10$^8$ $M_\odot$ (Zwaan {\it et al} 
~1997; Rosenberg \& Schneider 2002; Zwaan {\it et al} 2003) suffer severely
from small number statistics and from the systematics associated with
distance uncertainties and large scale inhomogeneities (Masters, Haynes 
\& Giovanelli 2004).  ALFALFA is specifically designed to survey enough solid angle to
detect several hundred objects with $M_{HI}$ $< 10^{7.5}$ and thus provide
a robust determination of the low HI mass slope of the HIMF. 
Already, ALFALFA has detected more galaxies with masses less than
$M_{HI}$ $< 10^{7.5}$ $M_\odot$ than included in all of the previous
HI blind surveys combined.

\begin{figure}[!t]
\plotone{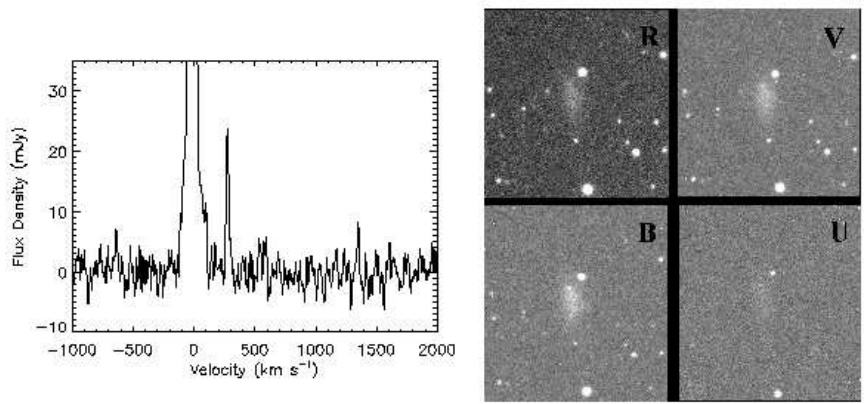}
\caption{Left: ALFALFA spectrum of AGC~112521 (HI0141+27). Right:
Broad-band images (courtesy of L. van Zee) using the WIYN 0.9m
telescope of the optical counterpart, a newly-discovered
low surface brightness dwarf member of the NGC~672 group.}\label{fig:haynesfig4}
\end{figure}

For her Ph.D. thesis, Am\'elie Saintonge 
(2007b) has undertaken a first study of some of the lowest 
HI mass galaxies detected early in the ALFALFA survey. 
All of them are nearby, optically faint and of low surface brightness. 
They exhibit a range of morphological clumpiness and rates of
massive star formation. Various
members of the ALFALFA consortium are pursuing multiwavelength
observations including broad-band and H$\alpha$ imaging, 
HI synthesis mapping and UV imaging with the GALEX satellite.
Of a first sample targeted for follow-up H$\alpha$ imaging,
15\% do not show the presence of HII regions.
While this work is still on-going, we present an illustrative
example of the kind of newly catalogued system that ALFALFA
finds. Figure \ref{fig:haynesfig4} shows the ALFALFA HI spectrum of
a new dwarf member of the NGC~672 group dubbed AGC~112521.
This object was first
detected in our ALFA-commissioning
precursor observations (Giovanelli {\it et al} 2005b) and recovered
in the early ALFALFA survey observations. As evident in Figure 
\ref{fig:haynesfig4}, the HI line emission from the galaxy is clearly detected
at c$z = +274$ km~s$^{-1}$, a narrow line full width at 50\% of the peak 
emission of 26 km~s$^{-1}$ and a total HI line flux density of 0.68
Jy-km~s$^{-1}$ (Saintonge {\it et al} 2007). Assuming its membership in the NGC~672
group at a distance of 7.2 Mpc,
the HI mass is $7.9 \times 10^6$ $M_\odot$. Its blue luminosity L$_B$ derived
from newly acquired images is similarly low: L$_B = 3.6  \times 10^6$ $L_\odot$,
so that $M_{HI}$/L$_B$ is 2.2. Adopting standard relations to convert multiband
magnitudes into stellar mass, we find that nearly half of the baryonic matter
is still in the form of gas.

With the promise of a final catalog 
of thousands of galaxies in the mass range $10^6 <$ $M_{HI}$ $< 10^8$ $M_\odot$,
the study of these gas rich systems is a prime science driver of ALFALFA.
Future ALFALFA studies will explore not only the low mass slope of the HI
mass function and its possible dependence on environment, but also
the distribution, morphology, star formation rate and chemical
enrichment history of these low mass gas-rich dwarfs.

\section{ALFALFA and the ``Void Problem''}\label{sec:haynes_voids}

As pointed out by Peebles (2001), numerical simulations based
on CDM predict that voids should contain large numbers of dwarf galaxies.
For example, the voids in the simulations of 
Gottl\"ober {\it et al} (2003) are criss-crossed by dark matter filaments
within which lie large numbers of very low amplitude inhomogeneities.
In fact, those authors predict that a void with a diameter of 20$h^{-1}$
Mpc should contain 1000 dark matter halos with masses of
10$^9$ $M_\odot$, and as many as 50 with masses ten times greater than that.
Peebles suggests that the failure to identify such a void population 
raises one of the principal challenges to CDM models.
Photoionization and baryonic blowout may suppress star formation or
perhaps the retention of any baryons within the low mass halos, but 
it still is not clear
that these processes are sufficient to explain the absence of galaxies 
in voids (Hoeft {\it et al}  ~2006).

Limits on the abundance and properties of void galaxies have been
placed by previous optical and radio studies.
Hoyle {\it et al} (2005) have used the SDSS dataset to show that
the OLF of void galaxies has a fainter break 
luminosity L$^*$ but a similar faint end slope to the overall 
SDSS luminosity function. In addition, they have shown that
void galaxies are typically blue, disk-like and have
high H$\alpha$ equivalent widths, making them excellent targets
for HI emission line surveys.
Previous surveys for HI in voids have exploited the VLA to conduct
blind HI surveys of the Pisces-Supercluster and its foreground void 
(Weinberg {\it et al} 1991; Szomoru {\it et al} 1994) and the
Bootes void (Szomoru {\it et al} 1993; 1996). 
Szomoru {\it et al} (1993) did find an isolated galaxy in Bootes, but its
mass of $5 \times 10^9$ $M_\odot$ and blue luminosity
L$_B$ of $-18.6$ exclude it as a true dwarf. In fact, those VLA surveys
sampled a relatively small volume and were hampered by poor spectral 
resolution (42 km~s$^{-1}$), not adequate to detect the lowest mass, narrowest
HI signals. 

Saintonge {\it et al} (2007) initiated a first, but limited, ALFALFA
analysis of galaxies in the void in front of the Pisces-Perseus supercluster 
at $cz \sim 2000$ km~s$^{-1}$. They detected no galaxies
in a large volume of 460 Mpc$^3$, whereas a scaling of the 
predictions of Gottl\"ober {\it et al} (2003) under the assumption 
that the dark-to-HI mass ratio is 10:1 predicts that ALFALFA
would have detected 38 HI sources. This very preliminary
result for a single void in only 2\% of the ALFALFA survey suggests that
the discrepancy between the predicted and observed abundance of
dwarf galaxies in voids cannot be reconciled by a population of
gas rich dwarfs. More credible results will be available when the
full ALFALFA
survey is completed.

\subsection{ALFALFA: Prelude to the SKA}
\label{sec:haynes_correl}

Although they acknowledge the volume and resolution limitations of the HIPASS catalog, 
recent papers have explored the behavior of the galaxy distribution it traces by
attempting to estimate its two--point correlation function $\xi (r)$. Meyer
{\it et al} (2007) conclude that gas-rich galaxies are among the most weakly
clustered galaxies known and suggest that the clustering scale length
$r_o$ depends strongly on rotational velocity, and thus, by implication,
on the halo mass. In contrast, Basilakos {\it et al} (2007)
argue that massive HIPASS galaxies show the same clustering properties
as optically-selected ones, but that
the low mass systems, $M_{HI}$ $< 10^9$ $M_\odot$, show a nearly uniform distribution.
Both of these studies are fraught with potential systematics because
of the shallow depth of HIPASS, the potential impact of source confusion
and the statistical limitations when 
its catalog is divided into subsets. As the follow-on to HIPASS,
ALFALFA is specifically designed to overcome many of the limitations of 
the earlier survey, thereby allowing a robust determination of the
HI-HI and HI-optical galaxy correlation functions and a quantitative study
of the biasing of the HI population relative to optical or IR-selected
samples and to the underlying density field. 

One of the prime science drivers of the SKA
is the undertaking of a billion galaxy redshift survey in the HI line
over the redshift range $0 < z < 2.5$ to explore the evolution
of the gas content of galaxies and constrain the dark energy equation
of state through the measurement of baryon acoustic oscillations
(Abdalla \& Rawlings 2004). Even allowing for the likely
increase in the gas content with $z$, only the most massive HI galaxies 
will be detected in emission at moderate redshift. As in the 
case of the low mass end of the HIMF, previous surveys have been
too shallow to detect the most massive HI galaxies. ALFALFA
has already detected more than twice as many massive galaxies with
$M_{HI}$ $> 10^{10.4}$ $M_\odot$ than all the previous HI blind surveys
combined. These massive galaxies exhibit a range of morphologies, colors and nuclear
concentration but all appear to be luminous disk systems. Many 
have stellar masses in the range corresponding to the ``transition mass''
($M_{stars} \sim 3 \times 10^{10}$ $M_\odot$) above which galaxies 
show a marked decrease
in their present- to past-averaged star formation rates (Kauffmann {\it et al} 
~2003). ALFALFA will
contribute its high mass detections to the GALEX--Arecibo--SDSS Survey 
(GASS; P.I.: D. Schiminovich), a new program to obtain measures
of the HI content at a gas mass fraction as low as 1.5\% of the
stellar mass in a sample of 1000 galaxies chosen by optical-UV
criteria to have stellar masses $> 10^{10}$ $M_\odot$. 
The combination of multiwavelength data will
provide new understanding of the physical processes that regulate
gas accretion and its conversion into stars in massive systems.
While ALFALFA$+$GASS will
characterize the properties of galaxies at $z \sim 0$, ambitious
future studies aimed at characterizing the evolution of galaxies
over the last 4 Gyr should be possible in the next few years
with Arecibo as well as the SKA precursor instruments. 

\section{Conclusions}\label{sec:haynes_concl}

ALFALFA is an ongoing survey with a detection catalog available in mid-2007
reflecting only about 15\% of the final survey.
Given its state, the full impact of ALFALFA is only beginning to
become evident, but the survey promises now to yield $> 25000$ extragalactic
HI detections when it is complete. It should yield robust measures of the
HIMF, the HI-HI and HI-optical correlation functions and their bias
parameters at $z = 0$, thereby establishing the present day constraints
on HI cosmology. The ALFALFA team is an open consortium and interested
parties are invited to follow the survey's progress via the ALFALFA website
{\it http://egg.astro.cornell.edu/alfalfa}.

\acknowledgements 
This work has been supported by NSF grants AST--0307661,
AST--0435697 and AST--0607007 and by the Brinson Foundation. 
The Arecibo Observatory is part of the National Astronomy and Ionosphere
Center which is operated by Cornell University under a cooperative
agreement with the National Science Foundation. ALFALFA is a team effort
and I acknowledge the significant and enthusiastic efforts of the
many individuals involved in its continuing success.


\end{document}